# A new way to constrain the densities of intragroup medium in groups of galaxies with convolutional neural networks

A. X. Shen* and K. Bekki*

*ICRAR M468 The University of Western Australia 35 Stirling Hwy, Crawley WA 6009, Australia*

**ABSTRACT**
Ram pressure (RP) can influence the evolution of cold gas content and star formation rates of galaxies. One of the key parameters for the strength of RP is the density of intragroup medium ($\rho_{\rm igm}$), which is difficult to estimate if the X-ray emission from it is too weak to be observed. We propose a new way to constrain $\rho_{\rm igm}$ through an application of convolutional neural networks (CNNs) to simulated gas density and kinematic maps galaxies under strong RP. We train CNNs using $9 \times 10^4$ 2D images of galaxies under various RP conditions, then validate performance with $10^4$ new test images. This new method can be applied to real observational data from ongoing WALLABY and SKA surveys to quickly obtain estimates of $\rho_{\rm igm}$. Simulated galaxy images have 1.0 kpc resolution, which is consistent with that expected from the future WALLABY survey. The trained CNN models predict the normalized IGM density, $\hat{\rho}_{\rm igm}$ where $0.0 \leq \hat{\rho}_{\rm igm,n} < 10.0$, accurately with root mean squared error values of 0.72, 0.83, and 0.74 for the density, kinematic, and joined 2D maps, respectively. Trained models are unable to predict the relative velocity of galaxies with respect to the IGM ($v_{\rm rel}$) precisely, and struggle to generalize for different RP conditions. We apply our CNNs to the observed H I column density map of NGC 1566 in the Dorado group to estimate its IGM density.

**Key words:** Galaxy: kinematics and dynamics – galaxies: clusters: intracluster medium – galaxies: disc.

## 1 INTRODUCTION

Ram pressure (RP) is an important physical process that affects galaxies in groups and clusters. It occurs when the gaseous halo of group and cluster galaxies interacts with the diffuse intracluster medium (ICM) of its host. The types of interactions that take place vary and include ram pressure stripping (RPS, Gunn & Gott 1972), external pressure on gas discs (Evrard 1991), thermal evaporation of the interstellar medium (ISM, Cowie & Songaila 1977), and viscous stripping of galaxy discs (Nulsen 1982). The interaction has an impact on many physical characteristics of the satellite galaxy, including significant changes to the star formation rate and galaxy morphology (Butcher & Oemler 1978; Fujita & Nagashima 1999; Bekki 2014).

RPS is heavily studied phenomenon and has been proposed as being a key mechanism for influencing galaxy evolution for a variety of galaxy types (e.g. Mori & Burkert 2000; Lucero, Young & van Gorkom 2005; Boselli & Gavazzi 2006; Mayer et al. 2006; Kawata & Mulchaey 2008; Yoon et al. 2017; Ramos-Martínez, Gómez & Ángeles 2018; Hausamman, Revaz & Jablonka 2019). RPS occurs when the ICM creates a wind that exerts a pressure to remove some or all of the galaxy's ISM. The analytical work by Gunn & Gott (1972) explores the mechanisms behind RPS, demonstrating that the extent of the stripping depends on the binding energy of the galaxy's ISM compared to the exerted RP. The result is various degrees of removal of the gas in the ISM, which can lead to drastic changes in star formation rates in spatial and temporal domains (e.g. Abadi,

Moore & Bower 1999; Balogh, Navarro & Morris 2000; Kenney, van Gorkom & Vollmer 2004; Cortese et al. 2012; Fossati et al. 2013; Safarzadeh & Abraham 2019; Tonnesen 2019).

The density of intragroup medium (IGM, $\rho_{\rm igm}$) and the relative velocity between a galaxy and the IGM ($V_{\rm r}$) are the two key parameters that determine the strength of RP. The relationship between RPS pressure $P_{\rm ram}$, relative velocity and IGM density is given by

$$P_{\rm ram} = \rho v^2 \tag{1}$$

as derived by Gunn & Gott (1972). It has been difficult to determine $\rho_{\rm igm}$ through direct observations of IGM, because IGM do not have high enough temperatures to emit *X*-ray.

Westmeier, Braun & Koribalski (2011) have attempted to constrain the $\rho_{\rm igm}$ by comparing the expected RP forces $P_{\rm ram}$ with gravitational forces $P_{\rm grav}$ stabilizing the disc for NGC 300. RPS of the disc gas occurs when the instability condition $P_{\rm ram} > P_{\rm grav}$ is fulfilled. The gravitational forces are determined by assuming the gravitational potential of the galaxy of the halo is dominated by the dark matter halo, neglecting the potential of the stellar and gas discs. Various $v_{\rm rel}$ values are compared for determining appropriate $\rho_{\rm igm}$ values under which RPS can occur. The method is not efficient and very time consuming, relying on high-resolution H I observational data and assumptions about the gravitational potential of the galaxies. Using CNNs to predict $\rho_{\rm igm}$ utilizes only density or kinematic maps of the disc galaxy undergoing RPS and is very quick to process, providing a practical method for constraining $\rho_{\rm igm}$ in observational data.

Bekki, Diaz & Stanley (2019) has shown that the 2D density maps of disc galaxies under RPS can have key information on the parameters of RPS. Bekki et al. (2019) applied machine learning, a subfield of computer science, using convolutional neural networks (CNNs) to constrain the 3D orbits of galaxies under RPS. The RPS can be determined by two angles ($\theta$ and $\phi$) that describe the direction of the 3D galaxy motion relative to the IGM. The CNN is trained using $10^4$ gaseous distribution images from RPS models with different $\theta$ and $\phi$, using cosine similarity as the performance metric (where $\cos \Theta = 1$ is a perfect prediction). The resulting average model performance on a unique test set of images was $\cos \Theta \approx 0.95$. Bekki et al. (2019) do not attempt to constrain $\rho_{igm}$ using CNNs.

*E-mail: austinxshen@gmail.com (AXS); kenji.bekki@uwa.edu.au (KB)

While the use of machine learning applied to astronomy is uncommon to date, there are various works that have also shown that CNNs can be applied effectively solve complex problems with high accuracy in the field. These include the categorizing of galaxy morphological types (Dieleman, Willett & Dambre 2015; Diaz et al. 2019), identifying shells and bubbles from turbulent molecular clouds (Van Oort et al. 2019), classifying radio images of extended sources (Aniyan & Thorat 2017). For the classification of S0 galaxy morphological types for instance, Diaz et al. (2019) are able to achieve accuracies exceeding 99 per cent with CNNs. CNNs have also been effective for various astronomical regression problems, such as the prediction of properties of the first galaxies from 2D images of 21-cm light-cones (Gillet et al. 2019), photometric galaxy profile modelling (Tuccillo et al. 2018) and estimation of galaxy cluster X-ray mass (Ntampaka et al. 2019).

The purpose of this paper is to apply machine learning with CNNs to predict RP parameters from simulated images of disc galaxies undergoing RPS. We attempt to constrain the IGM density, $\rho_{igm}$, the relative velocity of the disc galaxy with respect to the IGM, $v_{rel}$, and the RP, $P_{ram}$ directly. The images utilized are the 2D density and kinematics maps of cold gas in group member galaxies. A combination of the density and kinematic 2D maps (two-channel) will be used to compare the performance with isolated maps.

CNNs will be trained on $9 \times 10^5$ 2D images of gas density and kinematic maps from $9 \times 10^2$ unique simulations of disc galaxies undergoing RPS, with known values for the normalized IGM density and relative velocity. We generate 100 images with different viewing angle from each simulation from a constant time-step. The CNN model will be evaluated on separate sets of $10^4$ 2D maps from the same RPS galaxy simulations, different time-steps and different host environments to assess how well the model generalizes to new examples.

The plan of this paper is as follows. We describe the models utilized in Section 2, covering the RPS galaxy simulations and the training and architecture details of the CNN model adopted for prediction of RPS parameters. In Section 3, we present a summary of the information contained in 2D maps of gas distribution and kinematics and the prediction results for key RPS parameters $\rho_{igm}$ and $v_{rel}$ from the maps. In Section 4, we visualize CNN activations to better understand the important image features in the prediction task, discuss alternative implementations of the CNN model, and explore prediction performance in simulated data from different RP environments. The CNN is then applied to real observational images of H I density and the performance and limitations are discussed. In Section 5, we summarize the conclusions.

## 2 METHOD

### 2.1 Machine learning for image processing

Traditional approaches in machine learning for image processing utilize explicitly defined features (e.g. low-level features such as edge or corner detectors, or scale-invariant feature transform (SIFT) descriptors as seen in Smith & Brady 1997; Freeman, Pasztor & Carmichael 2000). These approaches tend to perform well in small data set regime, and learn from rulesets rather than from data. Deep learning, a field of machine learning, introduces a class of models that can learn from data rather than features. Such models have outperformed traditional machine learning methods in many problems, including image classification (e.g. Krizhevsky, Sutskever & Hinton 2012; Simonyan & Zisserman 2014) and object detection (e.g. Redmon & Farahadi 2018).

CNNs, introduced by LeCun et al. (1989), are a class of models in deep learning that provide a high-performing and end-to-end approach for various image processing problems (LeCun & Bengio 1995; LeCun et al. 1998; Krizhevsky et al. 2012). CNNs are adaptions to artificial neural networks (Hassoun 1995) that utilize convolutional filters for processing image data. CNNs are trained on a large labelled collection of raw images, and return outputs that can take various forms depending on the problem (e.g. class labels for image classification). For regression tasks such as ours, CNN models return a prediction of one or more continuous variables.

A large number of RPS disc galaxy models with different $\rho_{igm}$ and $v_{rel}$ are simulated in order to produce 2D density and kinematic maps of cold gas. The 2D maps of cold gas, of size $50 \times 50$ in pixels taken from galaxies with resolution of 0.7 kpc, are referred to as 'images' that are the inputs to train our CNN model. For each image, there are corresponding parameter values to predict: the normalized IGM density and relative velocity. Other parameters, such as direct prediction of $P_{ram}$, can be determined from $\rho_{igm}$ and $v_{rel}$. In addition to predicting RPS parameters independently with CNN models, we will also attempt simultaneous prediction ($\rho_{igm}$, $v_{rel}$) and other variables derived from the key RPS parameters (e.g. $P_{ram}$ and $v_{rel}^2$).

### 2.2 Disc galaxy

In our RPS galaxy simulations, we adopt the 'moving box model' for the evolution of gas in disc galaxies under strong RPS as seen in Bekki (2014). In this model, we first compute the orbit of a disc galaxy within its host group galaxy for a set of initial conditions using the adopted gravitational potential of the group. The strength of the RPS is estimated at each time-step for each model, calculated according to the position and velocity of the galaxy with respect to the group centre. Chemodynamical simulations provide spatial distributions of gas in disc galaxies at different time-steps under RPS.

The disc galaxies are composed of a dark matter halo, stellar disc, stellar bulge, and gaseous disc. In this study, we simulate luminous Milky-Way (MW) like disc models with $M_b = 10^{10} M_\odot$, $R_b = 3.5$ kpc, and $f_g = 0.1$. The mass ratio of the dark matter halo ($M_h$) to the disc ($M_s + M_g$) fixed at 16.7, and with $M_h = 10^{12} M_\odot$. We adopt the 'NFW' profile for the dark matter halo (Navarro, Frenk & White 1996) suggested from cold dark matter simulations, with the $c$-parameter value of 10 and virial radius of 245 kpc.

We vary the mass ($M_b$) and size ($R_b$) of the galactic bulge in our disc galaxy models explicitly to generate different maps. The gas mass fraction ($f_g = M_g/M_s$) is a additional free parameter in the galaxy simulations. The radial ($R$) and vertical ($Z$) density profiles of the stellar disc are proportional to $\exp(-R/R_0)$ with scale length $R_0 = 0.2 R_s$, and to $\mathrm{sech}^2(Z/Z_0)$ with scale length $Z_0 = 0.04 R_s$, respectively. The gas discs have size $R_g = R_s$ and radial and vertical scale lengths of $0.2 R_g$ and $0.02 R_g$ respectively. The disc of the present MW model has $R_s = 17.5$ kpc. The initial radial and azimuthal velocity dispersions are assigned according to epicyclic theory with

**Table 1** Description of the basic parameter values for the fiducial RPS model (T0) in a massive cluster of galaxies.

| Physical properties | Parameter values |
| --- | --- |
| Total halo mass (galaxy) | $M_{dm} = 1.0 \times 10^{12} \, M_\odot$ |
| DM structure (galaxy) | NFW profile |
| Galaxy virial radius (galaxy) | $R_{vir} = 245$ kpc |
| $c$ parameter of galaxy halo | $c = 10$ |
| Stellar disc mass | $M_s = 6.0 \times 10^{10} \, M_\odot$ |
| Stellar disc size | $R_s = 17.5$ kpc |
| Gas disc size | $R_g = 17.5$ kpc |
| Disc scale length | $R_0 = 3.5$ kpc |
| Gas fraction in a disc | $f_g = 0.1$ |
| Bulge mass | $M_b = 10^{10} \, M_\odot$ |
| Bulge size | $R_b = 3.5$ kpc |
| Mass resolution | $3.0 \times 10^4 \, M_\odot$ |
| Size resolution | 252 pc |

Toomre's parameter $Q = 1.5$. The parameters summarizing the disc galaxy RPS simulation are summarized in Table 1.

Various physical processes including star formation, chemical evolution, dust evolution, metallicity-dependent radiative cooling, feedback effects of supernovae, formation of molecular gas, are all included in this study. The details of the modelling of these processes are found in Bekki & Shioya (1998) and Bekki (2014, 2015). The Kennicutt–Schmidt law for galaxy-wide star formation (Kennicutt 1998) is adopted with gas density threshold for star formation at 1 atom cm$^{-3}$. The initial central metallicity of disc gas ([Fe/H]$_0$) and radial metallicity gradient are 0.34 and $-0.04$ dex kpc$^{-1}$ respectively.

The formation of molecular hydrogen from neutral on dust grains is modelled using the dust abundance of gas and the interstellar radiation field around the gas. Chemical yields for SNIa and SNII and those for asymptotic giant branch stars are adopted from Tsujimoto et al. (1995) and van den Hoek & Groenewegen (1997), respectively. The dust growth and destruction time-scales ($\tau_{acc}$ and $\tau_{dest}$, respectively) are set to be 0.25 and 0.5 Gyr, respectively. The canonical Salpeter initial mass function of stars (IMF) with the exponent of IMF being $-2.35$ is adopted.

### 2.3 Time-varying ram pressure force in the moving box model

In order to simulate RP forces, we consider a disc galaxy within its host group of galaxies to be embedded in hot ICM with temperature $T_{ICM}$, density $\rho_{ICM}$, and relative velocity (between ICM and disc galaxy) $V_r$. The ICM surrounding the disc galaxy is represented by smoothed particle hydrodynamics particles in a cube with the size $R_{box} = 3R_g$, where $R_g$ is the initial gas disc size corresponding to the stellar disc size in this work. This value of $3R_g$ is demonstrated to be large enough to model RPS in disc galaxies (Bekki 2014). This 'bound box model' is adopted in previous works (e.g. Abadi et al. 1999; Bekki 2014) so that the use of a large number of particles for representing the IGM in clusters of galaxies can be avoided. The galaxy is initially located at the centre of the cube Cartesian coordinate system with the direction of the orbit as along the $x$-axis.

Since we follow the orbit of the galaxy under the adopted cluster potential (constructed from the NFW profile), we can investigate both $\rho_{ICM}$ and velocity $V_r$ self-consistently at each time-step. Accordingly, we consider that the strength of RP force on the disc should be time-dependent and described by equation (1) where $\rho_{ICM}(t)$ and $V_r(t)$ are determined by 3D positions and velocities of a galaxy at each time-step in a simulation.

**Table 2** Description of the basic parameter values for the fiducial RPS model (T0) in a group of galaxies. The ICM mass is assumed to range from $0.015M_{dm}$ to $0.15M_{dm}$ in different models. The listed value is the maximum possible ($M_{ICM} = 0.15M_{dm}$).

| Physical properties | Parameter values |
| --- | --- |
| Total cluster mass | $M_{dm,g} = 1.0 \times 10^{13} \, M_\odot$ |
| Cluster virial radius | $R_{vir} = 0.56$ Mpc |
| $c$ parameter of cluster halo | $c = 6.0$ |
| ICM mass | $M_{icm} = 1.5 \times 10^{12} \, M_\odot$ |
| ICM temperature | $T_{icm} = 0.56 \times 10^9$ K |

The total mass of ICM within the cubic box is therefore time-dependent as follows:

$$M_{ICM}(t) = \rho_{ICM}(t) R_{box}^3. \quad (2)$$

As such, $M_{ICM}(t)$ is different from its initial value ($M_{ICM,0}$). Each ICM gas particle therefore requires that its mass ($m_{ICM}$) changes with time according to the change of $M_{ICM}$. For example, when a galaxy is approaching to the core of its host cluster, then $m_{ICM}$ can increase with time. We mainly investigate the group cluster model with $M_h = 10^{13} \, M_\odot$ and $T_{ICM} = 0.56 \times 10^9$ K because RPS is quite efficient in most galaxies close to the cluster core (Bekki 2014). These parameter values are summarized in Table 2.

In the following simulations, the spin axis of a disc galaxy under RPS is specified by two angles, $\theta$ and $\phi$. $\theta$ is the angle between the $z$-axis and the vector of the spin of a disc, while $\phi$ is the azimuthal angle measured from $x$-axis to the projection of the spin vector of a disc on the $xy$-plane. The direction of RP force with respect to gaseous motion in a local region of a galaxy depends strongly $\theta$ and $\phi$. This is a main reason why $\theta$ and $\phi$ can be inferred from the spatial distribution of gas influenced by RPS, as shown in Bekki et al. (2019).

The initial position of the disc galaxy is set to be $(x, y, z) = (R_i, 0, 0)$, where $R_i$ is the initial distance of the galaxy from the centre of its host cluster. $R_i$ is defined as follows:

$$R_i = f_p R_{vir} \quad (3)$$

where $f_p$ is a free parameter that controls the initial position and ranges from 0.1 and 0.5. The initial velocity of the galaxy is given by $(V_x, V_y, V_z) = (0, V_i, 0)$, where $V_i$ is as follows:

$$V_i = f_v v_c \quad (4)$$

where $v_c$ is the circular velocity of the galaxy at its initial position within the host cluster. The free parameter $f_v$ ranges from 0.5 to 0.7.

In this modelling of initial positions and velocities of a galaxy in a cluster, we consider that the gravitational potential of the cluster is spherical symmetric just for simplicity. The present simulations differ from those in Bekki (2014) in the sense that galaxies are initially within the virial radius of their host cluster. This is mainly because Bekki (2014) already found that RPS cannot strip the gas discs significantly until they become close to the inner regions of their cluster (see fig. 2 in Bekki 2014). Since the main purpose of this paper is to investigate the 2D density maps of galaxies under strong RPS, such modelling of galaxies (i.e. starting from strong RPS phases) would not be a problem.

### 2.4 Normalized $\rho_{igm}$ and $v_{rel}$ values

We set up the initial positions and velocities of galaxies (that control the minimum and maximum values of the two parameters for RP

before the start of simulations. Therefore, it is convenient for us to normalize these two parameters using these known minimum and maximum values. The values of the IGM density and relative velocity of the disc galaxy are normalized to range between 0 and 10 for convenience in training and prediction of the model.

We use $\hat{\rho}_{igm}$ and $\hat{v}_{rel}$ to denote the normalized values of $\rho_{igm}$ and $v_{rel}$, respectively. $\rho_{igm}$ is the dark matter density at the specific location within the galaxy group. $\rho_{igm}$ has lower and upper bounds defined by the dark matter halo, given by $\rho_{igm} = 0.015\rho_{DM}$ and $\rho_{igm} = 0.15\rho_{DM}$ (corresponding to the maximum possible density of the ICM) respectively. The relative velocity $v_{rel}$ values range from $0.3v_c$ to $0.7v_c$, where $v_c$ is the circular velocity at that location in the galaxy.

The following expressions are used to obtain normalized values from real values:

$$\hat{\rho}_{igm} = 10\left(\frac{\rho_{igm} - 0.015\rho_{dm}}{0.135\rho_{dm}}\right). \tag{5}$$

$$\hat{v}_{rel} = 10\left(\frac{v_{rel} - 0.3v_c}{0.7v_c}\right). \tag{6}$$

which converts the relative velocity and density from terms dependent on $v_c$ and $\rho_{dm}$ respectively into numerical values.

### 2.5 2D density and kinematic maps

In order to train the CNN we need to produce a larger number of 2D mass–density and (line-of-sight) velocity maps (often referred to as 'images') of simulated galaxies using the projected positions and the line-of-sight velocities ($V_{los}$) of gaseous particles in the galaxies. Each mass–density and velocity map generated has a corresponding RPS parameter ($\hat{\rho}_{igm}$ or $\hat{v}_{rel}$) that the CNN will attempt to predict based on the galaxy image features in the maps.

In order to derive the 2D density maps of simulated galaxies for $R \leq R_g$, we divide the gas disc ($R \leq R_g$) of a galaxy into $50 \times 50$ small areas (meshes) and estimate the mean gas density at each mesh point. The projected mass density of gas in a simulated galaxy can be estimated as follows:

$$\Sigma_{i,j,0} = \frac{1}{\Delta R_{i,j}^2} \sum_{k=1}^{N_{i,j}} m_k, \tag{7}$$

where $\Delta R_{i,j}$, $N_{i,j}$, and $m_k$ are the mesh size at the mesh point $(i, j)$, the total number of gas particles in the mesh, and the mass of a gas particle, respectively. In training a CNN, we use the logarithm of $\Sigma_{i,j,0}$ to base 10 as follows:

$$\Sigma_{i,j} = \log_{10} \Sigma_{i,j,0}. \tag{8}$$

The mesh size is $0.04R_g$ which corresponds roughly to 0.7 kpc for an MW-type disc galaxy. We also smooth out the density (velocity) field using a Gaussian kernel with the smoothing length ($h_{sm}$) of $0.05R_s$ (0.86 kpc). This smoothing is to mimic an observational resolution (e.g. beam size of a radio telescope) in a large survey of galaxies such as the WALLABY project. We discuss how the present results can depend on $h_{sm}$ in Section 4 later. We need to normalize the 2D data in order to feed the data into CNNs, and the normalized 2D gas density map can be derived as follows:

$$\Sigma'_{i,j} = \frac{\Sigma_{i,j} - \Sigma_{min}}{\Sigma_{max} - \Sigma_{min}}, \tag{9}$$

where $\Sigma_{min}$ and $\Sigma_{max}$ are the minimum and maximum values of $\Sigma$ among the $50 \times 50$ meshes in a model for a given projection. This normalization procedure is taken for each image at each time-step, and ensures that the 2D density ranges from 0 to 1. Therefore, the normalization factor is different in different models with different projections.

In order to generate a large number of density and kinematic maps, as typically required for training CNN models, we produce maps from 100 different, equally spaced, viewing angles for each RPS simulation. The data set used in our work consists of $1.1 \times 10^6$ unique maps for mass–density and line-of-sight velocity with corresponding $\rho_{igm}$ and $v_{rel}$ generated from $1.1 \times 10^4$ RPS simulations with varying initial conditions. $9 \times 10^4$ of these maps are used for training the CNN, while two distinct sets of $10^4$ maps are used for evaluating model performance. Variations on CNN inputs and prediction variables (e.g. two channel maps, or $v_{rel}^2$ prediction) that are used throughout this paper are derived from this data set.

### 2.6 Neural network architecture

CNNs have been used for a variety of image processing tasks (e.g. object detection, pose estimation, and image classification). The network architectures for these different tasks vary dramatically, but there are fundamental components that are common between these applications. The essential layers include a convolutional layer to extract a feature map from an image, an activation function that applies a non-linear transformation to the feature map, and a pooling stage to reduce the size of feature maps.

Convolutional layers in CNNs are responsible for feature point extraction from images (LeCun et al. 1989). Inputs to convolution layers are generally images, represented as multidimensional arrays of the image size (height and width in pixels) and colour in three channels. Generally, the colour channels are the separate red, green and blue intensities that compose the colour of a given pixel, but can also represent intensities for alternate colour spaces. For black and white images there is a single colour channel. The input images are convolved with a kernel, whose values are adapted by the learning algorithm to provide the optimal feature values for the learning task. These are passed through fully connected layers in order to predict the parameters of interest.

Pooling operations are used to reduce the resolution of feature maps by aggregating multiple features in a neighbourhood (Scherer, Müller & Behnke 2010). A pooled feature map is achieved by extracting a statistical summary of $n \times n$ patches across the input layer. The $n \times n$ patch is referred to as a pooling window, which can be of an arbitrary size and stride (can be overlapping). In max pooling, which is most commonly adopted in image processing problems, the value of each pooled feature map is the maximum value of the previous layer features in the $n \times n$ pooling window.

To enhance the performance of CNNs by avoiding overfitting, dropout is often used in the hidden layers of an artificial neural network (Hinton et al. 2012). Dropout works by randomly removing units of a layer of a neural network during training (with some probability, $p$). This introduces a regularization effect and causes each unit to become better at detecting features and independent of other units in the layer, which leads to an averaging effect across in the layer of the network and improves performance. Dropout values between $p = 0.4$ and 0.8 have been utilized in fully connected layers for various successful image processing CNN architectures, with $p = 0.5$ being most commonly adopted (Krizhevsky et al. 2012; Simonyan & Zisserman 2014; Szegedy et al. 2015).

The CNN adopted for predicting $\rho_{igm}$ consists of two convolutional layers ('Conv2D') with max pooling ('MaxPool') and a dropout layer, flattened to a dense layer, dropout, and a final dense output layer. The kernel adopted for the first and second Conv2D are of dimension ($3 \times 3 \times 32$) and ($3 \times 3 \times 64$) respectively, with $2 \times 2$ MaxPool layers. A dropout rate of $p = 0.25$ and $0.50$ is used for the first and second layers, respectively. The final fully connected layer connected flattened to 128 nodes before the output layer. Following each convolutional and fully connected layer, a ReLu activation function is utilized, with exception to output layer, where a linear activation is adopted. This architecture is similar to that used in Bekki et al. (2019), and is summarized in Fig. 1. The implementation of the CNN model described is performed using open-source Keras library; a high-level neural network API for deep learning. The CNN model architecture adopted for prediction of $\hat{\rho}_{igm}$ is the same as that used for the prediction of other parameters $\hat{v}_{rel}$ and $\hat{P}_{rps}$.

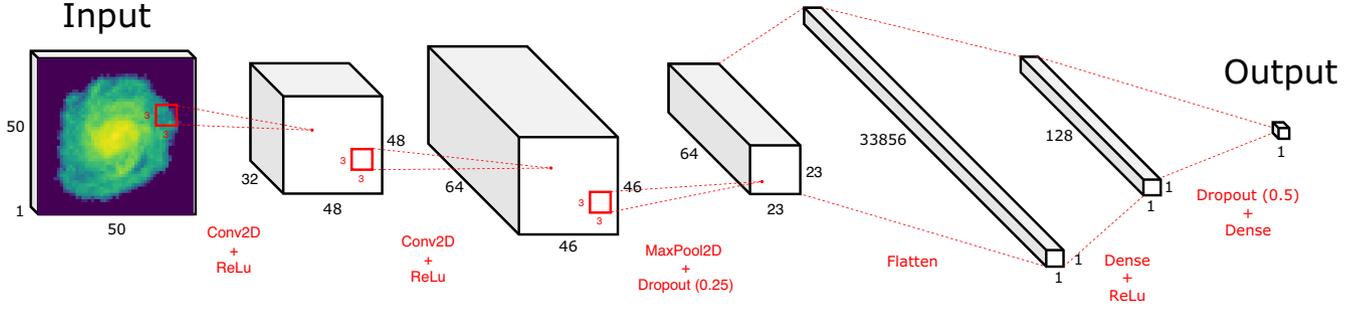

**Figure 1.** CNN architecture implemented for prediction of $v_{rel}$ or $\rho_{igm}$. The input image is a $50 \times 50$ 2D map of either the density or kinematics of the simulated disc galaxy. The model uses $3 \times 3$ kernels for the convolutional layers (of which there are two), and a $2 \times 2$ max-pooling layer to reduce the image size. The resulting tensor is flattened and connected to two dense layers that reduce the tensor shape to 128 units and 1 unit (output). The output in a single value for the estimate of $v_{rel}$ or $\rho_{igm}$ for the input image.

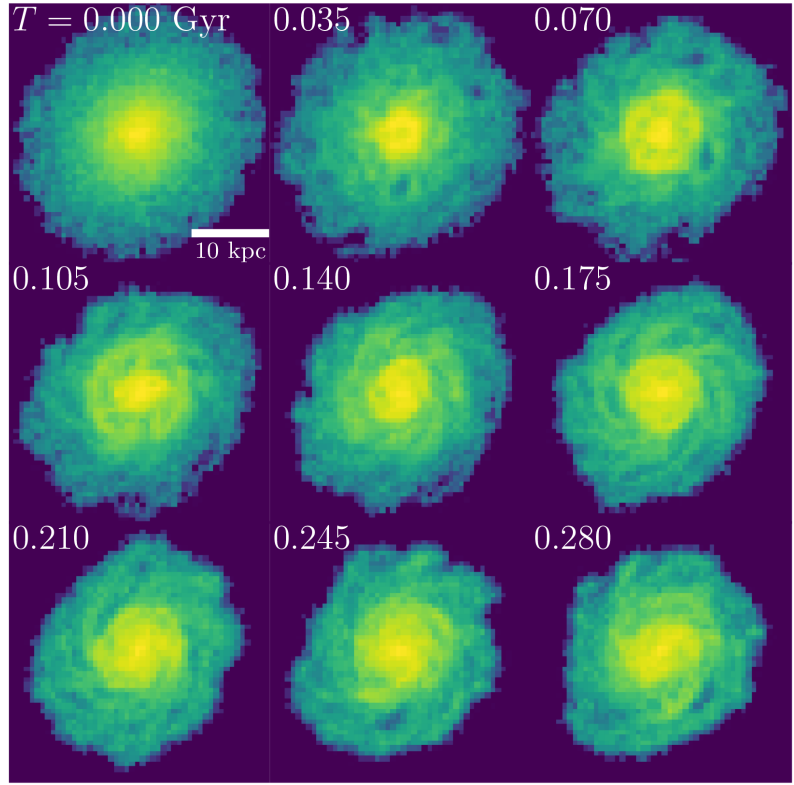

**Figure 2.** 2D density maps of a disc galaxy undergoing RPS with $\rho_{igm} = 0.15\rho_{dm}$ for simulation time-steps between $T = 0$ and $0.28$ Gyr. Changes to simulation time-step appear to have significant changes to density maps of the galaxy for early time-steps ($<0.14$ Gyr). The deformation in the density maps from RP appear to reduce and stabilize after $0.14$ Gyr.

### 2.7 Model training

There are three distinct data sets of simulated images used in the training and evaluation process. CNNs are trained on a 'training set' of $9 \times 10^4$ 2D density and kinematic maps. These images are generated from 100 different viewing angles of a disc galaxy from 900 unique RPS simulations with different conditions. A sample of the density and kinematic maps, with viewing angle $\theta = 45°$ and $\phi = 30°$ and varying time-steps, are shown in Figs 2 and 3 respectively. The CNN models are evaluated on a new set that are not used during the training phase, known as the 'validation set', comprising of $10^4$ additional images. The final performance of the CNN model is evaluated with another distinct 'test set' of $10^4$ images.

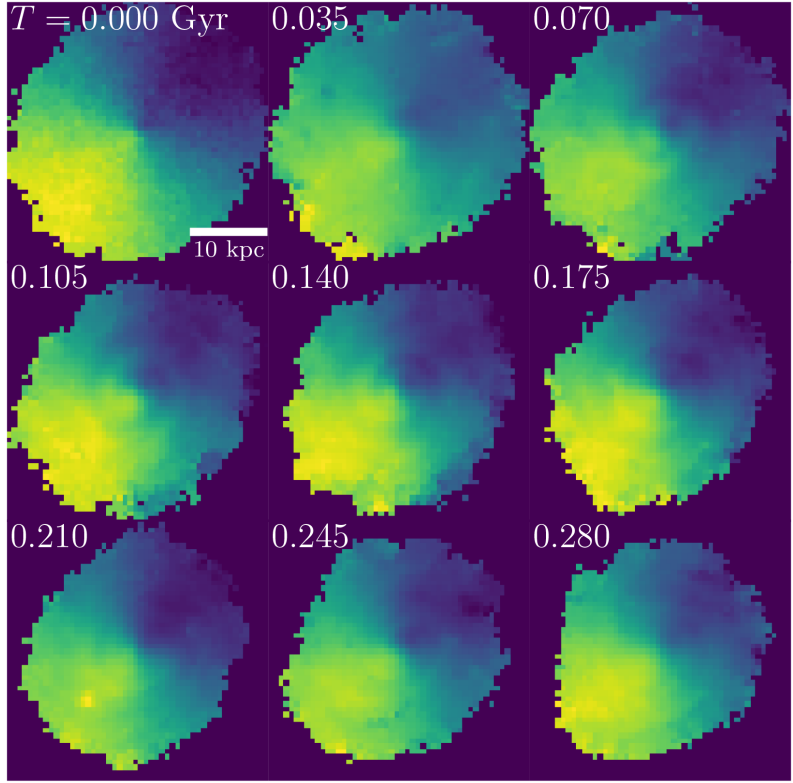

**Figure 3.** Time evolution of a 2D kinematic maps of a disc galaxy undergoing RPS at $\rho_{\rm igm} = 0.15\rho_{\rm dm}$ from $T = 0.0$ to 0.28 Gyr in 0.035 Gyr increments. Changes to kinematic maps are more subtle and random between time-steps, but large deformations appear to reduce after 0.14 Gyr.

The labels $\rho_{\rm igm}$ and $v_{\rm rel}$ are normaliZed to range between 0 and 10 for all images.

Each CNN model is trained over 300 iterations of the training data (epochs, denoted $N_{\rm epoch}$) with a batch size of 32. Training of the CNN model utilizes the "ADADELTA" (Zeiler 2012) learning rate method for gradient descent, with mean square error (MSE) as the optimisation metric. We explored training for 50 and 500 epochs, finding 50 to be insufficient based on the validation metrics and 500 to be excessive, yielding similar results to 300 epochs. The learning curve for 300 epochs in training $\hat{\rho}_{\rm igm}$ from 2D density maps is shown in Fig. 4. The model training rapidly improves the performance of the model, as measured by the MSE loss, prior to 100 epochs. Steady improvements are seen until 300 epochs, with the curve beginning to plateau after. This closely follows the CNN training methods observed in Bekki et al. (2019). Training duration for $9 \times 10^4$ images over 300 epochs is between 6.5 to 7 h, averaging 80 s per epoch utilizing a NVIDIA Tesla-K80 GPU with 8-core 32GB Intel Broadwell CPU.

The function on which the CNN model is evaluated is the root mean square error (RMSE) between the true IGM density from simulations and the model-predicted density. This is given by

$$\text{RMSE} = \sqrt{\frac{1}{n}\sum_{i=0}^{n}(\hat{\rho}_{\rm igm,p} - \hat{\rho}_{\rm igm,c})^2} \quad (10)$$

where subscript c denotes the value of the normalized density, subscript p denotes the predicted value, and $n$ is the number of training examples across which the loss is calculated. Low values of RMSE indicate good model performance, with RMSE = 0 indicating prediction and truth are equal. We consider RMSE values of model prediction against truth values less than 1.0 to be good predictions of RPS parameters. While RMSE is the primary metric for performance, we also evaluate the coefficient of determination ($R^2$) values for the prediction in order to determine the success of the model whilst accounting for variances in the data. We consider high values (>0.9) to be an indications that the trained model is performing well in the prediction task.

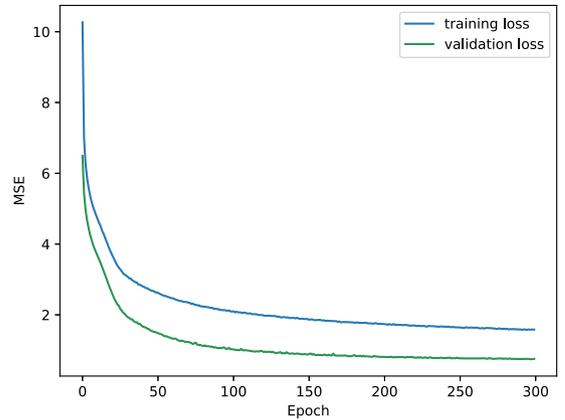

**Figure 4.** Learning curve to show how the loss (MSE) varies over 300 epochs for the training of our CNN to predict $\hat{\rho}_{\rm igm}$ from 2D density maps. Shows the training loss (blue) against the validation loss (green). Sharp improvements in the training and validation loss observed prior to 100 epochs, then steady improvement until 300 epochs.

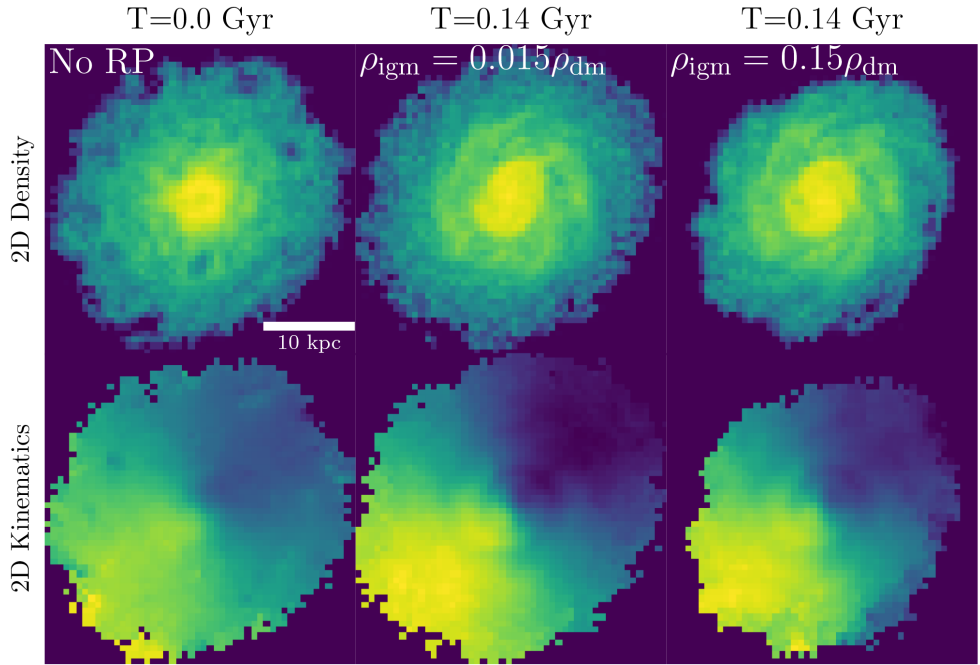

**Figure 5.** Comparison of the 2D density and kinematic maps of simulated disc galaxies that do not undergo RPS (leftmost images) with low RPS ($\rho_{igm} = 0.015\rho_{dm}$) and high RPS ($\rho_{igm} = 0.15\rho_{dm}$) for middle and right-hand columns, respectively. Reduced radial size and increasing elliptical shape with increasing RPS observed in both maps, with kinematic maps also showing higher noise and lower dispersion in line-of-sight velocity.

## 3 RESULTS

### 3.1 Influence of RPS on 2D density and kinematic maps

The effects of RPS on disc galaxies is visible in density and kinematic maps, and varies with the time-step of the simulation from which the maps are taken and the RPS parameters $\rho_{igm}$ and $v_{rel}$. Figs 2 and 3 show density and kinematic map images generated from different time-steps of a simulation. In these map images, points where the pixel value is zero represent areas of the simulation where there is no disc galaxy, so these points can be ignored. In each of the images, the viewing angle of the simulated disc galaxy is consistent with $\theta = 45°$ and $\phi = 30°$ for ideal comparison. By visualizing images of varying time-step and RP intensity, we are able to gain qualitative insights into the visual features that CNN models can learn from to predict RP parameters, and how they may change for different selected simulation time-step.

Fig. 2 shows how simulation time can significantly change the 2D density maps of cold ISM in disc galaxies in the fiducial model. The maps are taken at nine time-steps throughout the simulation, from $T = 0.0$ to $0.28$ Gyr, with increments of $0.035$ Gyr. In this evolution of 2D density maps we set $\rho_{igm} = 0.15\rho_{dm}$. Although early time-step images (<0.07 Gyr) already show the effects of RP, they are not well developed and change quickly with small increases to evolution time. The influence of RPS on the galaxy are best observed in later time-steps, with $T = 0.14$ Gyr or greater appearing sufficient for the effects of RPS on gas to be applied. A general pattern of increasing deformation and asymmetry with increasing time-step can be observed from the maps, though these changes appear to reduce above 0.14 Gyr.

We visualize the changes to kinematic maps from varying simulation times in Fig. 3. Again, we take $T = 0.0$–$0.28$ Gyr in 0.035 Gyr increments, producing nine kinematic maps of a galaxy experiencing consistent RP. A similar pattern is observed for simulation times below 0.14 Gyr; rapidly changing maps between time increments. The variations between time-steps is less apparent for kinematic maps, with greater noise in images between time-steps across the range of times. The large-scale changes to the kinematic maps appear to reduce after 0.14 Gyr (with exception for the $T = 0.21$ Gyr image) which is consistent with observations from density maps. The consistency in map image behaviour to show a majority of the extent of the galaxy deformation after $T = 0.14$ Gyr suggests that it is a reasonable time-step from which to take maps for prediction.

The effects of different RPS strength on the maps can be compared visually by choosing a constant time-step ($T = 0.14$ Gyr) and varying $\rho_{igm}$. In Fig. 5, we show three density and kinematic maps. The first image is taken from $T = 0.0$ Gyr, where no RP is experienced. In the second image, we take the map from $T = 0.14$ Gyr under weak RP using $\rho_{igm}$. In the third image, we show the maps under strong RP $\rho_{igm}$ at the same time-step. The observed changes in maps from weak to strong RPS are mainly in the radial size of the disc galaxy. Between low and high RP in both density and kinematic maps, the size of the galaxy is reduced and the shape of the galaxy becomes increasingly elliptical. In the kinematic map images, the difference in intensity of line-of-sight velocity on either side of the galaxy appears to reduce from low to high RP, but with less noise.

Fig. 6 shows a comparison of the pixel values for 2D density and kinematic maps between no RPS, weak and strong RPS ($\rho_{igm} = 0.015\rho_{dm}$ and $0.15\rho_{dm}$, respectively). The pixel values capture the amount of information in the images from which the model can learn, and allows us to quantify the differences in available information between no, weak and strong RPS galaxy images. In 2D density maps, the pixel values represent the line-of-sight density of the gas. In 2D kinematic maps pixel values represent the average velocity magnitude of the particles along that line of sight. In both cases, the values are normalized to range between 0 and 1. In our comparison of pixel value distributions we remove zero values as they represent empty background that is not part of the disc galaxy.

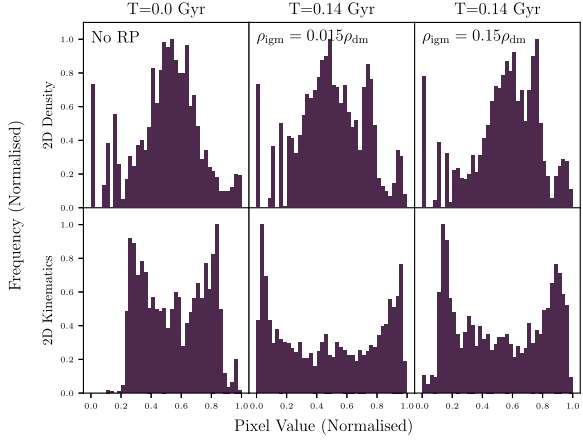

**Figure 6.** Comparison of the 2D density and kinematic map pixel value distributions of simulated disc galaxies. For 2D kinematic maps, pixel value represents the average velocity. For 2D density maps, the pixel value represents the average density of that area of the galaxy along the line of sight. Comparison of three cases: disc galaxies experiencing no RPS (leftmost images), low RPS (middle), and high RPS (right). Strength of the RPS is determined by $\rho_{igm}$ in these simulations.

**Table 3.** Summary of RMSE and $R^2$ values for prediction of our trained CNN models on test data against expected true values for different 2D maps. Low values of RMSE below 1.0, and high $R^2$ above 0.9 are an indication of good performance in prediction. By this measure, we are able to predict $\rho_{igm}$ from 2D maps of density, kinematics, and from joined maps successfully.

| 2D map | Variable | RMSE | $R^2$ |
|---|---|---|---|
| Density | $\rho_{igm}$ | 0.72 | 0.929 |
| | $v_{rel}$ | 2.23 | 0.373 |
| | $v_{rel}^2$ | 2.22 | 0.369 |
| | $P_{ram}$ | 1.05 | 0.686 |
| | $(\rho_{igm}, v_{rel})$ | 1.66 | |
| Kinematics | $\rho_{igm}$ | 0.83 | 0.907 |
| | $v_{rel}$ | 2.38 | 0.283 |
| Joined | $\rho_{igm}$ | 0.74 | 0.925 |
| | $v_{rel}$ | 2.25 | 0.360 |

In Fig. 6, we observe that the pixel value distribution in 2D density maps are observably skewed between different RPS strengths. A left-skewing distribution is observable in the high $\rho_{igm}$ case while a right-skewing distribution is observed for lower or moderate $\rho_{igm}$. There is also a greater number of higher density points for high $\rho_{igm}$ as expected in stronger RPS simulations. In 2D kinematic maps, there are observable differences between galaxies that experience no RPS and those that experience RPS of some strength. Pixel value differences between weak and strong RPS 2D kinematic maps however are not immediately observable.

### 3.2 Constrained for $\rho_{igm}$

The trained CNN is consistently able to successfully predict $\hat{\rho}_{igm}$ from 2D density maps, 2D kinematic maps, and joined two-channel maps with a high accuracy. We are able to predict $\hat{\rho}_{igm}$ with RMSE values of 0.72, 0.83, and 0.74 for the density, kinematic and joined maps, respectively. Similarly, the $R^2$ values are 0.929, 0.907, and 0.925. These model evaluation scores, along with those for the prediction of other RPS parameters, are summarized in Table 3. High values of $R^2$ also indicate that our model is successful in predicting $\hat{\rho}_{igm}$ from the different maps.

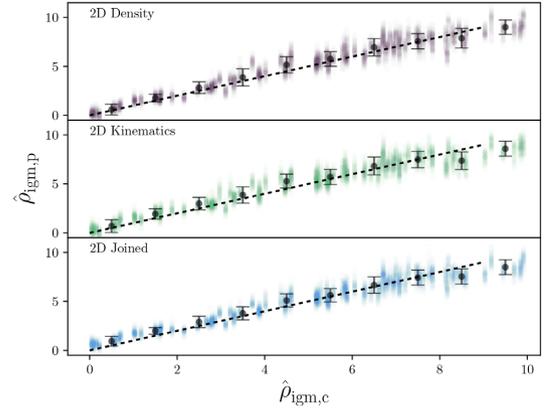

**Figure 7.** Comparison of CNN density prediction results ($\hat{\rho}_{igm,p}$) against truth values ($\hat{\rho}_{igm,c}$) for 2D density and kinematic maps independently, and joined in a two-channel image. Dashed line in each image indicates perfect prediction of the parameter (RMSE = 0). Error bars show one standard deviation of the predicted values in bins of width 1.0. Small deviation around ideal prediction line indicates successful prediction of $\rho_{igm}$ with CNN model.

Fig. 7 shows scatter plots of the CNN predicted $\hat{\rho}_{igm}$ (vertical axis) against the true $\hat{\rho}_{igm}$ values (horizontal axis) from the simulation for each example in the test data set for each input map type. Here, we use the subscript p to denote the predicted values from the CNN model, and c to denote the correct or truth values from the simulation. Error bars represent one standard deviation in the prediction values for each bin of 1.0 width. The dotted black line in each of the subfigures indicates an ideal prediction of $\hat{\rho}_{igm}$. In each of the subfigures, the ideal prediction line falls within one standard deviation of the CNN predictions, with the exception of very high values of $\hat{\rho}_{igm}$ (9.0−10.0). Given the flat distribution of $\rho_{igm}$ values in the training data, we consider this relatively poor performance $\hat{\rho}_{igm}$ to be more likely attributed to fewer examples in the validation and test data sets. The small deviations around ideal prediction demonstrate that the CNN model is able to accurately predict $\hat{\rho}_{igm}$ from density, kinematic, and joined map images.

### 3.3 Constrained for $v_{rel}$

Our trained model for prediction of $\hat{v}_{rel}$ by the trained CNN model is unsuccessful. With the same model architecture, number of training examples and training hyperparameters the RMSE values for prediction on the test set are 2.23, 2.38, and 2.25 for the density map, kinematic map, and joined images, respectively. The $R^2$ values are 0.373, 0.283, and 0.360 for the three different maps, respectively. Both RMSE and $R^2$ metrics are very low, indicating that the model has no predictive capability for constraining $v_{rel}$.

An additional CNN model is trained to attempt to predict a normalized value for $\hat{v}_{rel}^2$. For this task, only the 2D density maps were used to train and evaluate the model. This attempts to provide the model with a simpler prediction task, since $\hat{v}_{rel}^2$ is proportional to the RPS pressure that disturbs gas the disc galaxy density maps based on equation (1). Performance using $\hat{v}_{igm}^2$ is similarly poor compared to that for $\hat{v}_{\rm rel}$, with RMSE = 2.22 and $R^2 = 0.369$ on the 2D density test data set.

### 3.4 Simultaneous prediction of $\rho_{\rm igm}$ and $v_{\rm rel}$

By adapting the shape of the CNN model to have two output nodes rather than one, we are able to attempt the simultaneous prediction of $\rho_{\rm igm}$ and $v_{\rm rel}$ in a single inference. This simultaneous prediction approach has been found to be effective in various applications of deep learning to astronomy problems (Gupta et al. 2018; Schmit & Pritchard 2018; Gillet et al. 2019; Hassan, Andrianomena & Doughty 2020). Aside from having two output units required for simultaneous prediction, all aspects of the CNN architecture are unchanged. The training procedure and hyperparameters are also unchanged. For this task, we use only 2D density maps for training and evaluation of the performance given their success in other tasks.

We find prediction of ($\rho_{\rm igm}$, $v_{\rm rel}$) concurrently from 2D density maps to be unsuccessful. The RMSE for prediction on the test set is 1.66, which is below our threshold for successful prediction. The $R^2$ value cannot be evaluated for this task. It is interesting to note that this result is better than the prediction of $v_{\rm rel}$ but worse than that for $\rho_{\rm igm}$ alone.

We can explore the effectiveness of simultaneous prediction further by evaluating the CNN model predictions of $v_{\rm rel}$ and $\rho_{\rm igm}$ separately, and determining RMSE and $R^2$ scores for each independently. The model is able to learn to predict $\rho_{\rm igm}$ successfully, with RMSE = 0.82 and $R^2 = 0.909$ on the test set. The performance for $v_{\rm rel}$ however is poor, with RMSE = 2.20 and $R^2 = 0.385$. This result explains why the RMSE is between that for predicting $\rho_{\rm igm}$ or $v_{\rm rel}$ independently, and is consistent with the results found for the other trained CNN models in predicting RPS parameters.

### 3.5 Constrained for $P_{\rm ram}$

We also attempt to predict the strength of the RP directly, using a normalized value determined by $P_{\rm ram} = \rho_{\rm igm} v_{\rm rel}^2$. Normalized values of $\rho_{\rm igm}$ and $v_{\rm rel}$ are used to compute $P_{\rm ram}$, which is then normalized to range between $0 \le \hat{P}_{\rm ram} < 10$. In this task, we use only 2D density maps for training and testing the model.

Performance in predicting RP strength on our test image set is worse than that of $\hat{\rho}_{\rm igm}$, but better than prediction of $\hat{v}_{\rm rel}$ with RMSE = 1.05 and $R^2 = 0.686$. Fig. 8 shows the predictions against truth values for $\hat{P}_{\rm ram}$. Each point is a prediction-truth pair, with the dashed black line indicating perfect prediction and black error bars indicating the standard deviation for bins of width 1.0. Interestingly, a reasonable performance is observed (low scatter around dashed line) at low normalized $\hat{P}_{\rm ram}$ values when compared to higher values. This pattern is likely related to the successful predictions of $\rho_{\rm igm}$, which contributes greater to $P_{\rm ram}$ at low values, compared to the relatively performance of $\hat{v}_{\rm rel}$. As $P_{\rm ram}$ increases the contribution from the $v_{\rm rel}^2$ term dominates, which we have been unsuccessful in predicting in our attempts. Consequently, the model poorly predicts $P_{\rm ram}$ in this regime.

The trained CNN models are able to make successful predictions of $\hat{\rho}_{\rm igm}$ from density, kinematic and joined maps, but are not successful for prediction of $\hat{v}_{\rm rel}$ or $\hat{P}_{\rm ram}$ from any of the different variations of map images. All training jobs utilized the same architecture (with the exception of joined maps where the input channels are adapted) and use the same optimizer and training hyperparameters. The summary of prediction performance for $\hat{\rho}_{\rm igm}$, $\hat{v}_{\rm rel}$, and $\hat{P}_{\rm ram}$ by different trained

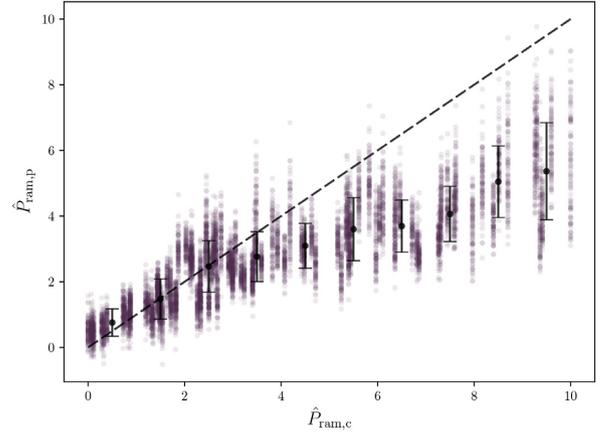

**Figure 8.** Prediction of $\hat{P}_{\rm ram}$ using 2D density images by trained CNN model on test set data. Ideal performance (prediction equal to truth) given by the dashed black line. Black error bars show the standard deviation of points in bins of width 1.0. Small standard deviation at low values of $\hat{P}_{\rm ram}$, indicative of good performance, likely related to the accurate prediction of $\hat{\rho}_{\rm igm}$ dominating during weak RPS (true values tend fall within a standard deviation of the average predictions for weak RPS). Larger scatter at high $\hat{P}_{\rm ram}$, indicative of poor performance, is possibly explained by dominance of $v_{\rm rel}$ at high $P_{\rm ram}$ and CNN model inability to successfully predict $v_{\rm rel}$.

CNN models applied to image sets of 2D density, kinematic, and joined maps is captured in Table 3.

## 4 DISCUSSION

### 4.1 Visualizing learned features

A common criticism of CNN models is the lack of interpretability around the underlying features learned by the convolutional layers used to make predictions. A summary of the various methods explored for better understanding the features learned by CNN models can be found in Qin et al. (2018). Visualizing the activation values (the output of the Conv2D and ReLu block in this instance) of early layers during the feed-forward process for a given CNN inference task is a simple method for interpreting image features used for prediction. Inspection of activations in our CNN models allows us to compare the important features for the model to our intuition about the expected deformations of galaxies under RPS.

To visualize the activations, we take the output from the second convolutional block of an inference. The resulting tensor has shape (64, 46, 46, 1) as shown in Fig. 1. We take the mean of the activation values through the 64 channels for each pixel coordinate in order to reduce this tensor to an image of shape (46, 46, 1). The resulting 'feature map' can be compared directly between different trained CNN models and for different input maps to the network at inference. Figs 9 and 10 show the feature maps corresponding to the activations for the trained CNN model in predicting $\hat{\rho}_{\rm igm}$ from 2D density and kinematic map inputs, respectively. In visualizing these activation maps we gain a better understanding of the features learned for prediction tasks from 2D density and kinematic map images.

For each prediction task we compare the feature maps for low, moderate, and high values of $\rho_{\rm igm}$. Given the relative success in the prediction of $\rho_{\rm igm}$ compared to $v_{\rm rel}$ we look only at feature maps for tasks in predicting $\rho_{\rm igm}$. This provides insight into the features that are learned from 2D density and kinematic maps, which are likely used for other prediction tasks with the same input. We normalize the feature map by dividing all activation values by the highest activation value across the three feature maps. The activation values are normalized across the feature maps in order to allow for visual comparison of the both the relative magnitude and location.

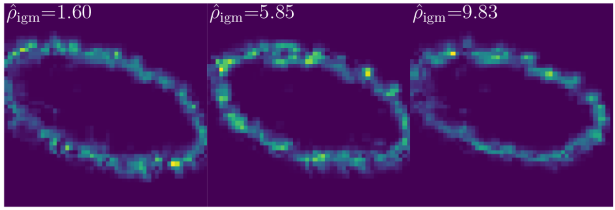

**Figure 9.** Mean activation maps (normalized) for the second layer outputs of CNN model inference with 2D density map images in prediction of $\hat{\rho}_{\rm igm}$. Values of $\hat{\rho}_{\rm igm}$ are 1.60, 5.85, and 9.83 for the left-hand, middle, and right-hand maps, respectively. In each feature map the edge of the disc is activated, providing an indication of the useful features of 2D density map images in RPS parameter prediction tasks. Although no significant trend emerges, there are slightly higher activations observed for higher values of $\hat{\rho}_{\rm igm}$ (particularly noticeable in the middle image). The location of activated points on the map images also appear to change, with higher activation points at the top of the disc in the moderate and high compared to the low $\hat{\rho}_{\rm igm}$ images.

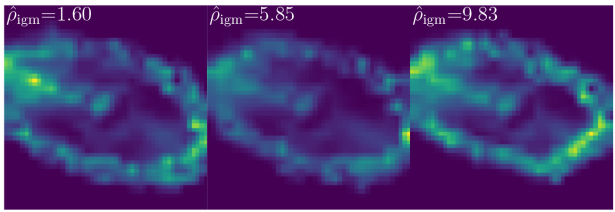

**Figure 10.** Mean activation maps (normalized) for the second layer outputs of CNN model inference with 2D kinematic map images in prediction of varying $\hat{\rho}_{\rm igm}$. Values of $\hat{\rho}_{\rm igm}$ are 1.60, 5.85, and 9.83 for the left-hand, middle and right-hand maps, respectively. More of the galaxy image is activated for kinematic maps compared to density maps, with the edge of the disc and the centre being activated in each feature map shown. While no significant patterns can be observed for in the activations for kinematic maps, activation values for the rightmost feature map are noticeably greater. Middle feature map shows unusually low activation values.

Fig. 9 shows three activation maps for prediction of weak, moderate, and strong $\rho_{\rm igm}$ from 2D density maps, corresponding to RPS parameters ($\hat{\rho}_{\rm igm}$, $\hat{v}_{\rm rel}$) are (1.60, 4.85), (5.85, 5.14), and (9.83, 4.18) for the left-hand, middle, and right-hand maps, respectively. In selecting the examples to inspect, we have attempted to keen $v_{\rm rel}$ constant to isolate its effect on the image. The feature maps show increased activations at the edge of the disc galaxy, which is consistent with our expectations for 2D density maps. There appears to be greater activations on the middle and right-hand images (higher $\hat{\rho}_{\rm igm}$) when compared to the left-hand image (low $\hat{\rho}_{\rm igm}$). Other than the locations of high-activation value pixels in the centre and right-hand images, there are few significant observable differences.

Kronberger et al. (2008) explore the effects of RPS on internal kinematics of spiral galaxies through *N*-body hydrodynamical simulations. They find that for edge-on RP the velocity fields are increasingly asymmetric, the effects of the interaction appear in both inner and outer parts of the disc, and that the kinematics are highly dependent on direction of the acting RP. The 2D kinematic maps used for our CNN training and prediction contain less information than the 2D velocity fields explored in Kronberger et al. (2008), as direction of the particles are not shown in our maps. Consequently, it is difficult to determine which image features will be useful for prediction of RPS parameters for kinematic maps.

Fig. 10 shows the mean activation maps for the second layer of the network in a forward pass with 2D kinematic map images in the prediction of varying values of $\hat{\rho}_{\rm igm}$. The RPS parameter values in the figure are the same as those in Fig. 9. In observing these kinematic maps, we find that there are more areas of the image that are activated for a prediction task. The edge of the disc galaxy still appears to provide useful feature information, but unlike the 2D density maps the centre of the galaxy in kinematic maps also appear to contribute to the prediction. It is difficult to determine clear trends in image features for increasing RPS parameters. It is clear that the mean activations are higher for high $\hat{\rho}_{\rm igm}$ compared to low $\hat{\rho}_{\rm igm}$, though moderate values appear to show even lower mean activations.

### 4.2 Varying simulated image time-step

The disc galaxy map images and corresponding RPS parameters are taken from the same evolution time ($T = 0.14$ Gyr) in our simulations. Galaxies undergoing similar RPS can appear different depending on the time-step from which the images are generated, as can be shown in Figs 2 and 3. As such, we can expect predictions of $\rho_{\rm igm}$ and $v_{\rm rel}$ from a trained CNN model, which relies on visual features, to differ for maps generated from different time-steps. In real survey images, the environmental conditions of the RPS are not constrained by the duration over which the galaxy has experienced RP effects. Therefore, in order for our trained model to be effective at constraining RP parameters for real images from surveys the predictions require to be accurate (low RMSE) for prediction of maps generated from different time-steps.

We use a small set of 2D density maps to test how well our CNN model, which has been trained only for a specific time-step, generalizes to other time-step values. In this data set, we predict and vary only $\hat{\rho}_{\rm igm}$ across time-steps ranging from $T = 0.14$ to 0.28 Gyr in 0.035 Gyr increments. Images with $T < 0.14$ Gyr are ignored as they are unlikely to show the extent of the RPS effects in that environment. We produce images from three different $\rho_{\rm igm}$ values: $\rho_{\rm igm} = 0.15\rho_{\rm dm}$, $0.015\rho_{\rm dm}$, and $0.045\rho_{\rm dm}$, which correspond to $\hat{\rho}_{\rm igm} = 10.0$, 0.0, and 3.0, respectively. The 2D density and kinematic maps of each time-step for $\rho_{\rm igm} = 0.15\rho_{\rm dm}$ are shown in Figs 2 and 3.

Fig. 11 shows the prediction of $\hat{\rho}_{\rm igm}$ for three constant values ($\hat{\rho}_{\rm igm} = 10.0$, 0.0, and 3.0 for purple, green, and blue curves, respectively) from map images taken from $T = 0.14$ to 0.28 Gyr in 0.035 Gyr increments. The dashed horizontal lines in the figure indicate the expected $\hat{\rho}_{\rm igm}$ values, while the transparent lines indicate the least-squares linear regression curve for the points. It is interesting to show that the predicted value at $T = 0.14$ Gyr, from which our maps are sampled, tend to be lower than the correct value. Predictions from greater time-steps ($T = 0.21$ Gyr for $0.015\rho_{\rm dm}$ and $0.045\rho_{\rm dm}$, and $T = 0.24$ Gyr for $0.15\rho_{\rm dm}$) are closer to the true $\hat{\rho}_{\rm igm}$ value. Given the unbiased predictions of test data, as shown in Fig. 7, this pattern is more likely an attribute of the sample selected (e.g. specific viewing angle) than an inherent flaw in the trained CNN model.

In each value of $\hat{\rho}_{\rm igm}$, we see that the predictions vary slightly from the expected value. The predicted value of $\hat{\rho}_{\rm igm}$ tend to increase with increasing simulation time-step. This is a strong indication that our model is not able to generalize effectively to data from different time-steps. Given our model assumes the training data are representative of the population, and increasing exposure to RPS environments will cause the disturbance of disc gas to increase, it is reasonable to see increased prediction values for increasing time-steps. In order to train a better model that generalizes better to different time-steps, we require simulated data that shows the evolution of the 2D density and kinematic maps across different time-steps and for different $\rho_{igm}$ and $v_{rel}$. Additional training data will allow the time-step to be explicitly captured in the model prediction, and will be important future work to ensure that our model can be applied to real data.

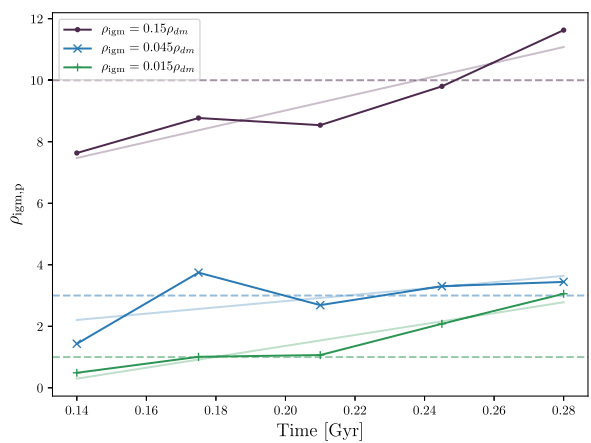

**Figure 11.** Predictions of $\rho_{igm}$ for three RPS simulations under different conditions ($\rho_{igm} = 0.15\rho_{dm}$, $0.045\rho_{dm}$, and $0.015\rho_{dm}$ for the purple, blue, and green curves, respectively) from 2D density maps over multiple time-steps ranging from 0.14 to 0.28 Gyr. For low values of $\rho_{igm}$, the predictions across different time-steps appears relatively constant but in the correct regime. For the high values of $\rho_{igm}$ low time-step 2D density images result in poor prediction, with predicted values increasing with time-step. Trained model does not extend well for images generated from a different time-step.

### 4.3 Gas mass fraction dependence

Since the disc restoring force against RPS of gas depends on the gas surface density for a given stellar gas density (e.g. Gunn & Gott 1972), the models with very high $f_g$ cannot be so strongly influenced by RP compared to the fiducial model. On the other hand, the models with low $f_g$ can be influenced more heavily under the same RP than models with high $f_g$. Therefore, our prediction can be inaccurate if we do not also account for these differences in $f_g$. Here, we test the effect of CNN models that are constrained with a constant $f_g$ on test case density and kinematic maps where $f_g$ varies. In future works, we will extend the training data of our CNN models to include different values for $f_g$.

We generate new images of 2D density maps from simulations with different initial gas mass fractions to determine if the model is able to generalize to new data. We generate $3 \times 10^3$ 2D density maps from simulations with $f_g = 0.03$, 0.455, and 0.091, and all other parameters unchanged. Although the two models with very low (0.03) and high (0.455) $f_g$ are extreme cases, they can be stringent tests for our CNN.

The trained CNN model makes reasonable predictions for $f_g = 0.091$ but fails where $f_g = 0.455$ and 0.03. The RMSE values for the prediction are 4.71, 2.47, and 1.68 for $f_g = 0.03$, 0.455 and $f_g = 0.091$ respectively. Although the performance of the model is relatively poor in each of these cases, the reasons for poor performance may differ in each case. For $f_g = 0.03$, the predictions appear to be consistently greater than the real values with the exception of a few outliers at high $\hat{\rho}_{igm}$. Where $f_g = 0.455$, there appears to be no pattern in the prediction of $\hat{\rho}_{igm}$. For the case where $\rho_{igm} = 0.091$ the predicted values appear closer to true values despite relatively high RMSE with the exception of outliers.

**Table 4.** Summary of the repeated convolutional blocks used for the alternative CNN architecture models. In each alternate architecture, convolutional layers utilize $3 \times 3$ kernel sizes and max-pooling layers use $2 \times 2$ stride.

| Alternate 1 | Alternate 2 | Alternate 3 |
| --- | --- | --- |
| conv2D ($k = 3$) | conv2D ($k = 3$) | conv2D ($k = 3$) |
| maxPool2D ($s = 2$) | conv2D ($k = 3$) | conv2D ($k = 3$) |
| | maxPool2D ($s = 2$) | maxPool2D ($s = 2$) |
| | | batchNorm |
| | | Dropout ($p = 0.5$) |

### 4.4 Alternative model architectures

Selecting the ideal CNN model architecture for an image processing problem can be a difficult task due to the many possible permutations of layers, activation functions, and hyperparameters values. We compare our chosen model architecture for predicting $\hat{\rho}_{igm}$ from $50 \times 50$ 2D density maps (see Fig. 2) with alternate architectures that take inspiration from high-performing image classification models AlexNet (Krizhevsky et al. 2012) and VGG16 (Simonyan & Zisserman 2014). The architectures described in these works are much more complex and deeper than our selected model, owing to their need to reduce input higher resolution input images and predict more parameters (one prediction parameter for each classification). As such, rather than using the architectures described exactly for our prediction task, we instead use the different repeated patterns of layers seen in these other architectures (sometimes referred to as convolutional 'blocks') and adapt the shape so they are appropriate for our task.

To reduce the input maps into features we explore the use of three different convolutional blocks. In the CNN model, the input will passed through three sequential convolutional blocks, which are then flattened to a feature vector and mapped through dense layers to a single prediction value. The convolutional blocks utilized are summarized in Table 4. The complexity of the CNN architecture, as measured by the number of trainable parameters, increases from Alternate 1 to Alternate 3. We also train a model with the same architecture as our original model without dropout layers in the network to assess the impact of dropout on predictive performance. The performance of each of these trained models are compared to the original architecture.

In our comparison, we keep all other layers and hyperparameters aside from the convolutional blocks consistent across alternate architectures. Following the application of three convolutional blocks, we flatten the resulting tensor into a ($1 \times 1 \times N$) feature vector, where $N$ varies for each architecture. The feature vector is connected to a sequence of dense and dropout ($p = 0.5$) hidden layers (128 and 64 units, respectively) before mapping to a single output unit. A ReLu activation is used after each hidden layer. The hyperparameters [number of epochs (300), optimizer (AdaDelta), loss function (MSE), learning rate (0.01), and batch size (32)] used for training remain unchanged from those used for the training of our original model.

The performance of each network in predicting $\hat{\rho}_{igm}$ from density maps on a test and validation data set is summarized in Table 5 using RMSE as the evaluation metric. In each model, we are able to successfully predict $\hat{\rho}_{igm}$ from 2D density maps. Our original architecture performs slightly better compared to each of the alternative architectures on the validation and test data sets. This is likely a result of using hyperparameters that are optimized for the training of a different model. In applying fixed hyperparameters to the new training jobs, the new models do not benefit from fine-tuning performed during the model training process of our original architecture. As such, it would be reasonable to expect worse performance without the deliberate selection of optimizer, number of epochs and learning rate for the specific architecture. However, given the similar success in prediction performance for all models compared, is is likely that large performance gains can not be achieved with small architectural changes and that the architecture adopted is sensible for our prediction task.

**Table 5.** Summary of prediction RMSE of alternate CNN model architectures for $\hat{\rho}_{igm}$ from 2D density maps compared to original model on test and validation data sets. Each alternate CNN architecture is able to successfully predict $\hat{\rho}_{igm}$ to high accuracy but not as precisely as our original architecture. Model without dropout performs better than alternate architectures, but not as well as our original architecture with dropout.

| Architecture | Validation RMSE | Test RMSE |
| --- | --- | --- |
| Original | 0.78 | 0.72 |
| Alternate 1 | 1.01 | 0.94 |
| Alternate 2 | 0.95 | 0.90 |
| Alternate 3 | 0.97 | 0.98 |
| No dropout | 0.87 | 0.81 |

### 4.5 Prediction against NGC 1566

In our future works we will try to apply the newly developed CNNs to real images of galaxies in groups obtained from ongoing large H I surveys, such as the Widefield ASKAP L-band Legacy All-sky Blind surveY (WALLABY: Koribalski et al. 2020) and future surveys by the Square Kilometre Array (SKA). Since we do not have a large sample of H I column density maps, we test our CNN by applying it against an H I image of NGC 1566 obtained in the WALLABY survey (Elagali et al. 2019). This Dorado group is ideal for testing our CNNs because the group mass estimated by the velocity dispersion of group member galaxies is roughly $10^{13}$ M$_{\odot}$ (Elagali et al. 2019), which is consistent with the adopted group mass in this study. In order to apply the CNN, we divide the H I image into $50 \times 50$ regions (exactly the same as the number of pixels used in the simulated images) and thereby run an inference on the image with the trained CNN. Fig. 12 shows the 2D map of H I column density obtained by Elagali et al. (2019).

Our CNN predicts $\hat{\rho}_{igm} = 2.063$, which implies that the gas disc of this spiral disc galaxy is being influenced by RP in this group. Elagali et al. (2019) find that the asymmetry in the structure and kinematics of the H I disc and thus suggested that RPS is ongoing in this galaxy. If this group has a mass of $10^{13}$ M$_{\odot}$ like our group model, the estimated value of $\hat{\rho}_{igm} = 2.063$, which is equivalent to $\rho_{igm} = 0.0429 \rho_{dm}$, implies that the total IGM mass is only 4.3 per cent of the dark matter mass. However, since the total mass of this group is yet to be determined precisely (Elagali et al. 2019), we cannot make a robust conclusion of this. If the WALLABY survey reveals the 2D H I maps for many galaxies in groups with estimated total mass, then our new method will be able to be applied to these images to estimate the masses and densities of the IGM.

## 5 CONCLUSION

We have investigated a large number of models for disc galaxies under RPS using hydrodynamical simulations in order to produce a large number of 2D density and kinematic maps of simulated galaxies. We have then used $9 \times 10^4$ images to train a CNN to predict IGM density, $\rho_{igm}$, and relative velocity, $v_{rel}$ and RP, $P_{ram}$ of the simulated galaxies. The CNN model utilizes the same architecture for each prediction task using the RMSE on $10^4$ test images as the evaluation metric for performance.

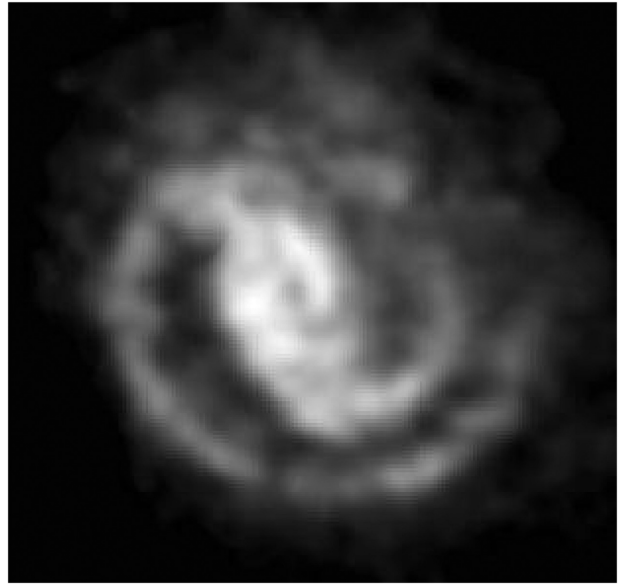

**Figure 12.** A $50 \times 50$ image of H I column density for spiral galaxy NGC 1566 in the Dorado group by Elagali et al. (2019).

The principle conclusions are as follows:

(i) A CNN that can accurately predict $\rho_{igm}$ can be trained with RMSE of 0.72, 0.83, and 0.74 for a test set of 2D density, kinematics, and joined maps, respectively. Inspection of hidden layers of the CNN model reveal high activations at the border of disc galaxies for 2D density maps, suggesting high values of $\rho_{igm}$ are associated with disc deformations in the simulated images.

(ii) CNNs are not able to predict $v_{rel}$ with high accuracy with either density or kinematic maps. The RMSE of these predictions is significantly greater than that for $\rho_{igm}$, taking 2.23, 2.38, and 2.25 for the test 2D density, kinematics and joined maps, respectively. An attempt to predict $v_{rel}^2$ using test density maps is also unsuccessful with RMSE = 2.22. Density map prediction hidden layers show no patterns between activation regions in the image and $v_{rel}$.

(iii) Prediction of $P_{ram}$ is relatively unsuccessful for 2D density maps, with test performance RMSE = 1.05. The weak $P_{ram}$ regime dominated by $\rho_{igm}$ can be predicted reasonably, but as $P_{ram}$ increases to larger values dominated by $v_{rel}$ prediction becomes unsuccessful. The result is consistent with our attempts at predicting $\rho_{igm}$ and $v_{rel}$ independently.

(iv) In simultaneous prediction of ($\rho_{igm}$, $v_{rel}$) from 2D density maps the model performs poorly, with test RMSE = 1.66. This is attributed entirely to the unsuccessful prediction of $v_{rel}$ in the joined prediction task, which when assessed independently is shown to have a test RMSE = 2.20 and $R^2 = 0.385$. Prediction for $\rho_{igm}$ is successful with test RMSE = 0.82 and $R^2 = 0.909$.

(v) Alternate CNN model architectures with more convolutional layers and blocks do not increase the performance in predicting $\rho_{igm}$ from 2D density maps. The original model used in Bekki et al. (2019) outperforms three other CNN models used with RMSE = 0.72 on the test map set (compared to 0.94, 0.90, and 0.98 for other models). The model performs slightly worse with dropout layers removed (RMSE = 0.81).

Although we will need to improve the accuracy of the CNN-based prediction RPS parameters in our future works using different CNN architectures and larger data sets, we suggest that the presented new method is promising because it will allow us to estimate $\rho_{\rm igm}$ for a large number of galaxies in groups simultaneously. Furthermore, this method can be applied for distant groups at high redshift $z$, where the X-ray emission can be too weak to estimate the properties of IGM of groups, if future SKA observations reveal the 2D density and H I maps of the member galaxies with sufficient resolution. In our future studies, we intend to extend this analysis to various environments including high-$z$ groups and clusters.

## ACKNOWLEDGEMENTS

The authors are grateful to the referee for a careful reading of the manuscript and constructive comments that improved this paper.

## DATA AVAILABILITY

The data underlying this article will be shared on reasonable request to the corresponding author.